\pdfoutput=1 
\documentclass [10pt] {article}
\usepackage[usenames]{color}

\usepackage [utf8] {inputenc}
\usepackage [T1] {fontenc}

\usepackage {epsfig}
\usepackage {fullpage}

\usepackage {url}
\usepackage {alltt}

\usepackage{amssymb}
\usepackage{amsmath}

\usepackage{listings}

\newcommand {\institution} {{Mines-ParisTech (CEMEF)}}
\newcommand {\corresponding} {{Correspondence~: {bernard.monasse@mines-paristech.fr}}}

\newcommand{\gmol}[1] {{{#1} \: {g/mol}}}
\newcommand{\kcalMolAngsquare}[1] {{{#1} \: kcal/mol/\angstrom^2}}
\newcommand{\kgMolNmsquare}[1] {{{#1} \: kg/mol/nm^2}}

\newcommand{\fs}[1] {{#1}\:\mathit{fs}}
\newcommand{\ns}[1] {{#1}\:\mathit{ns}}
\newcommand{\ps}[1] {{#1}\:\mathit{ps}}

\newcommand{\nm}[1] {{#1}\:\mathit{nm}}

\newcommand{\degree}[1] {{#1}^{o}}

\newcommand{\temp}[1] {{{#1}\:K}}

\newcommand{\angstrom} {{\text {\normalfont\AA}}}

\newenvironment{myfigure}
{\begin{figure}[!!htb]
\begin{center}}
{\end{center}
\end{figure}}

\newcommand{\myfig}[1] {{Fig.\:{#1}}}

\newcommand{\gauche} {{\it gauche}}
\newcommand{\gauchep} {{\it gauche'}}
\newcommand{\trans} {{\it trans}}

\newcommand{\alkane}[2] {{C_\mathit {#1}H_\mathit {#2}}}

\newcommand{\CC}[2] {{C_{#1}C_{#2}}}
\newcommand{\CH}[2] {{C_{#1}H_{#2}}}

\newcommand{\CCC}[3] {{C_{#1}C_{#2}C_{#3}}}
\newcommand{\CCH}[3] {{C_{#1}C_{#2}H_{#3}}}
\newcommand{\HCC}[3] {{H_{#1}C_{#2}C_{#3}}}
\newcommand{\HCH}[3] {{H_{#1}C_{#2}H_{#3}}}

\newcommand{\CCCC}[4] {{C_{#1}C_{#2}C_{#3}C_{#4}}}
\newcommand{\CCCH}[4] {{C_{#1}C_{#2}C_{#3}H_{#4}}}
\newcommand{\HCCC}[4] {{H_{#1}C_{#2}C_{#3}C_{#4}}}
\newcommand{\HCCH}[4] {{H_{#1}C_{#2}C_{#3}H_{#4}}}

\newcommand{\corr}[1] {{\bf  {#1}}}

\newcommand{\equi}[1] {{\bf eq}_{{#1}}}

\newcommand{\paramcos}[2] {{P_{{#1}_{#2}}}}

\newcommand{\EtE} {{\it EtE}}
\newcommand{\FFT} {{FFT}}

\title{\bf A Correlation Method for Deriving  UA Intra-Molecular
Potentials from AA Molecular Dynamics Simulations:
Application to Alkanes}

\newcommand {\fbsignature} 
{{\sc Fr\'ed\'eric Boussinot} \\
\institution}

\newcommand {\bmsignature} 
{{\sc Bernard Monasse} \\
\institution}

\author {\bmsignature\footnote {\corresponding} \and \fbsignature}

\begin{document}
\maketitle

\begin{abstract}

  Unified-Atom (UA) force fields are usually constructed using a Boltzmann-inverse
  method based on distributions obtained from Monte-Carlo simulations.
 A new method of constructing UA force fields from All-Atom (AA) molecular dynamics
 simulations is proposed. In this method, one determines time correlations between
 oscillators of the same type: between a CC bond and the connected CH bonds;
 between a valence angle CCC and all valence angles sharing at least one hydrogen with it;
 between a dihedral angle CCCC and the dihedral angles sharing the two central carbons with it.
 In the case of no correlation between lengths or angles, energies of the oscillators
 are independent. In this case, the AA and UA components of the force fields are identical.
 When a correlation is total, the UA potential of an object is the sum of the energies
 of the oscillators coupled to it. Partial correlations are also possible.
 Several kinematic matching tests of the AA molecule with the corresponding
 UA molecule are considered: mean periods of vibration of the oscillators;
 energies associated with presence probabilities; mean end-to-end lengths of the molecules.
 The method is, as an example, applied to aliphatic molecules,
 simulated with the OPLS-AA force field.

\end{abstract}


\section {Introduction \label {section:introduction}}
Molecular Dynamics \cite {MolecularDynamics} (MD) simulations at the All-Atom (AA) level are a recognized technique for predicting the behaviour of molecules at the atomic scale. The time scale is limited by a
time-step of no more than 1 femto-second ($\fs 1$) for molecules with CH bonds ($\approx \fs {10}$). Thus, an AA simulation up to one nano-second ($\ns 1$) requires at least $10^{6}$ time-steps.
A large number of simulations, dealing for example with diffusion phenomena, require longer times.

Incorporating hydrogen atoms into {\it grains} is the standard technique for bypassing the time limitation associated with the time-step.  This approach, called {\it Unified-Atom} (UA), also reduces the number of degrees of freedom and, therefore, the complexity of the models.  In an UA hydrocarbon chain, for example, each grain is composed of one carbon atom and two or three hydrogen atoms (three hydrogens at both ends of the chain) and the mass of a grain is the sum of the mass of all the atoms it contains.  The UA level is actually the first step of the {\it Coarse-Grained} (CG) approach in which grains can have more complex structures.

UA simulations should have kinematics and dynamics as close as possible to those in AA, under the same conditions.  This proximity must exist both locally, at the level of each oscillator, and globally, with regard to the conformation of the molecule.  As the kinematics and dynamics of the molecules depend on the potentials, these must be coherent at the AA and UA scales.

AA atoms and UA grains are structured within molecules by bonds, valence angles and dihedral angles which are oscillators whose behaviors are specified by their potentials grouped in force fields.  To the intra-molecular potentials of the force fields are added inter-molecular potentials, corresponding to van der Waals interactions.  The AA force field that will be considered in this text is OPLS-AA \cite{OPLS}.  In this one, the potentials of the bonds and of the valence angles are harmonic functions, those of the dihedral angles are {\it triple-cosine} functions, and the inter-molecular potentials are {\it Lennard-Jones 6-12} functions.

Most of the UA force fields are determined using Monte-Carlo simulations performed on dense matter at AA level \cite{OPLS-UA, OPLS-UA1, OPLS-UA2, OPLS-UA3, OPLS-UA4, AMBER-UA, CHARMM-UA, GROMOS-UA}.  They are extracted from the simulation data using a Boltzmann-inverse approach, providing energies from the probabilities of presence of the simulation (micro-)states.  These determinations are centered on dihedral angle potentials and inter-molecular potentials.  Bond potentials and valence angle potentials are generally defined, without any particular justification, as being the same as in AA.

It is recognised that the terms of intra-molecular potential energy are much more important than those of inter-molecular interaction energy \cite{WUNDERLICH}.  In a crystal, in particular, the equilibrium conformation of an isolated molecule is only slightly modified by inter-molecular interactions \cite{NATTA} and thus depends mainly on intra-molecular components.  In this paper, only the intra-molecular components of the OPLS-UA force field \cite{OPLS-UA4} are considered.

In the proposed approach, intra-molecular UA potentials are defined from simulations of individual AA molecules, in the vacuum and at different temperatures.  Each oscillator (bond, valence angle, dihedral angle) is studied individually.  This study is based on the determination of temporal correlations between oscillators of the same type and the UA potentials are deduced from the correlations found.

\section {Correlation Method\label{method}}

The correlation method is a general method, to deduce a UA potential
using data from Molecular Dynamics AA simulations. It is presented
here in the framework of alkanes which are linear (aliphatic)
hydrocarbon chains.  To simplify the analysis, only fragments of
molecules will be considered in which each carbon always has two
associated hydrogens (and not three for the extreme carbons). Thus,
the fragments considered will always have a structure of the form
$\alkane {n}{2n}$.  At the UA level, fragments are chains
of grains $G_{n}$. A natural correspondence exists between the AA
fragment $\alkane {n}{2n}$ and the UA fragment $G_{n}$ which
associates the grain $G_i$ to the carbon $C_i$ and to its two linked
hydrogens (the center of the UA grain being the carbon atom).

We number the atoms in the $\alkane {N}{2N}$ fragment in the following
way: the carbons are numbered consecutively, from 0 to $\mathit {N-1}$
and the two carbon-bonded hydrogens $C_n$ are numbered
$\mathit {N+2n}$ and $\mathit {N+2n+1}$.  
Figure \ref{C6H12Structure} shows the numbering of the atoms of 
fragment $\alkane {6}{12}$ as well as that of the grains of the
corresponding fragment $G_6$. In the following, to simplify, we will
no longer speak of fragment but only of a molecule.
\begin{myfigure}
\begin{center}
  \includegraphics [height=4cm] {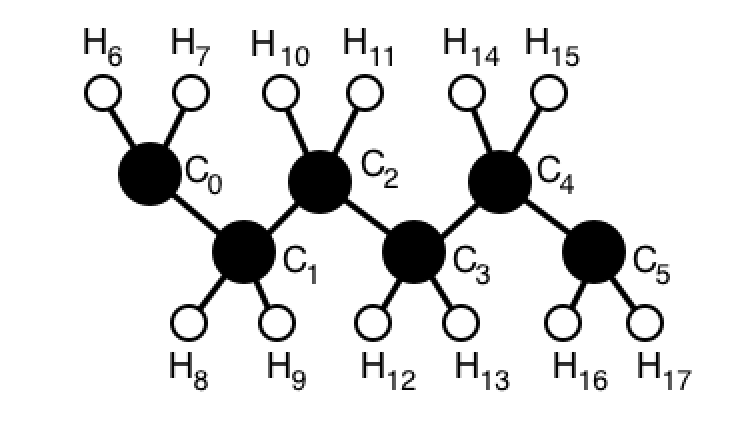}
  ~~~
\includegraphics [height=4cm] {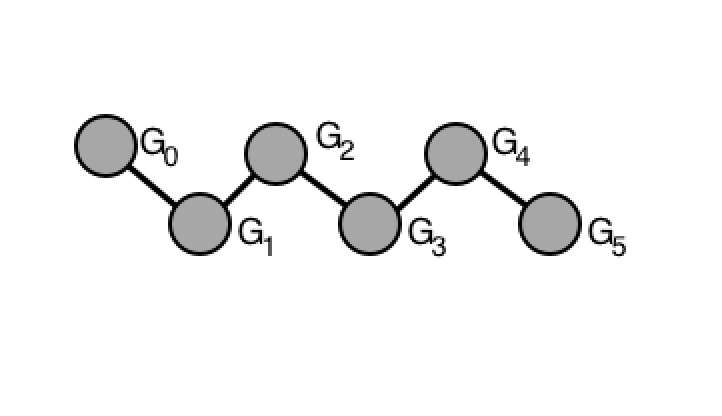}  
\caption{\small {\bf Left Image:}
Numbering of alkane fragment $\alkane {6}{12}$ atoms.
  {\bf Right Image:} UA corresponding fragment $G_6$.
}
\label{C6H12Structure}
\end{center}
\end{myfigure}

The temporal correlation method analyses the average evolution of an
oscillator of the carbon chain according to those of the
hydrocarbon oscillators of the same type (bond, valence angle or
dihedral angle) that are coupled to it.  Three types of couplings are
defined:
\begin{itemize}
\item[$\bullet$] 
Couplings between CC and CH bonds.  The CH bonds coupled to the
  bond $\CC {i}{i+1}$ in the molecule $\alkane n {2n}$ are the 4
  bonds $\CH {i}{n+2i}$, $\CH {i}{n+2i+1}$, $\CH {i+1}{n+2i+2}$ and
  $\CH {i+1}{n+2i+3}$.

\item [$\bullet$]
Couplings between CCC valence angles and 
  valence angles CCH or HCH. The
  CCH angles coupled to the $\CCC {i}{i+1}{i+2}$ angle in the molecule
  $\alkane n {2n}$ are the four angles $\HCC{n+2i}{i}{i+1}$,
  $\HCC{n+2i+1}{i}{i+1}$, $\HCC{i+1}{i+2}{n+2i+4}$ and
  $\CCH{i+1}{i+2}{n+2i+5}$. The only HCH angle coupled with
  $\CCC {i}{i+1}{i+2}$ is the angle $\HCH {n+2i+2} {i+1} {n+2i+3}$.
  
\item [$\bullet$]
Couplings between CCCC dihedral angles and 
  dihedral angles CCCH or HCCH. The
  HCCH angles coupled to $\CCCC {i}{i+1}{i+2}{i+3}$ in the molecule
  $\alkane n {2n}$ are the four dihedral angles
  $\HCCH {n+2i+2} {i+1}{i+2} {n+2i+4}$,
  $\HCCH {n+2i+3} {i+1}{i+2} {n+2i+4}$,
  $\HCCH {n+2i+2} {i+1}{i+2} {n+2i+5}$ and
  $\HCCH {n+2i+3} {i+1} {i+2} {n+2i+5}$.  The CCCH angles coupled with
  $\CCCC {i}{i+1}{i+2}{i+3}$ in the molecule $\alkane n {2n}$ are
  the four angles $\HCCC {n+2i+2} {i+1}{i+2} {i+3}$,
  $\HCCC {n+2i+3} {i+1}{i+2} {i+3}$, $\HCCC {i} {i+1}{i+2} {n+2i+4}$
  as well as the angle $\CCCH {i} {i+1} {i+2} {n+2i+5}$.

\end{itemize}

Molecular dynamics providing the coordinates of all atoms at each
time-step, one can measure the length of the bonds and the angular
value of the valence angles and of the dihedral angles.

For each pair of coupled oscillators, the pairs formed by the values
of the hydrocarbon oscillator and those of the carbonated oscillator,
are measured at the same time, form a cloud of points.  This cloud results
from the fluctuation over time of the exchanges of kinetic energy
between the two oscillators.  We analyze the statistical dependence
between pairs of values by means of a linear regression using a least
square method that allows to get a correlation factor.  There are
three possible scenarios, with different consequences for the
determination of the UA potential:

\begin{itemize}

\item [$\bullet$]
  The slope of the linear regression line is zero. This
  means that there is no temporal correlation between the value of the
  of the carbonated oscillator and that of the hydrocarbon
  oscillator.  The carbonated oscillator thus evolves on average
  independently from the hydrocarbon oscillator.  In the absence of
  correlation, the vibration of an AA component does not depend on
  those of the coupled oscillators. The potential energy of the UA
  component therefore does not include any energy contribution from
  hydrocarbon coupled oscillators.  The UA component behaving in the
  same way as the corresponding AA carboned component, they must have
  the same potential.

\item [$\bullet$]
  The correlation factor is equal to $-1$ or $1$.  The points are
  distributed on either side of the mean regression line, due to the
  exchange of kinetic energy between these oscillators.  The
  correlation is then perfect and the values of the hydrocarbon
  oscillator are in mean directly related to those of the carbonated
  oscillator.  Therefore, the potential energy of the corresponding UA
  component must be the sum of the potential energy of the AA
  carbonated component, and of the potential energies of the
  hydrocarbon coupled oscillators.

\item [$\bullet$]
  The correlation factor is between $-1$ and $1$ and
  is different from $0$.  Such a case is characteristic of a
  correlation between the hydrocarbon oscillators and the carbonated
  component.  In this case, the UA and AA potentials are different.
  The potential energy of the UA oscillator comes from the AA
  potential energy of the carbonated oscillator and from a portion of
  those of the coupled hydrocarbon oscillators.

\end {itemize}

\paragraph {Notations.}
In AA as in UA, the bond potentials are harmonic:
\begin{equation}
{\cal U}_{\alpha} (l) = k_{\alpha} (l - \equi {\alpha})^2 
\end{equation}
where ${\cal U}_{\alpha}$ is the potential of the oscillator of type
$\alpha$, $\equi {\alpha}$ the equilibrium length of the oscillator,
$k_{\alpha}$ its stiffness and $l$ a length.  The various possible
cases for $\alpha$ are, in AA, the CC and CH bonds and in UA, the GG
bond.

Similarly, in AA as in UA, the valence angle potentials are
harmonic:
\begin{equation}
{\cal U}_{\alpha} (\theta) = k_{\alpha} (\theta - \equi {\alpha})^2 
\end{equation}
where ${\cal U}_{\alpha}$ is the potential of the oscillator of type
$\alpha$, $\equi {\alpha}$ the equilibrium angle of the oscillator,
$k_{\alpha}$ its stiffness and $\theta$ an angle. The
various possible cases for $\alpha$ are, in AA, CCC, CCH, HCH and in
UA, GGG.

The potentials of the AA and UA dihedral angles, on the other hand,
have a triple-cosine shape \cite{OPLS}:
\begin{equation}
{\cal U}_\alpha (\phi) = (
   \paramcos{\alpha}{1}   (1 + cos (  \phi))
+ \paramcos{\alpha}{2}  (1 - cos (2\phi))
+ \paramcos{\alpha}{3}  (1 + cos (3\phi))
) / 2 \label{equation:triple-cosine}
\end{equation}
The various possible cases for $\alpha$ are in AA, CCCC, CCCH, HCCH
and in UA, GGGG. All dihedral angles are ranging from $\degree {0}$ to
$\degree {360}$.  It is important to note that the harmonic potentials
have a single minimum, while hydrocarbonat potentials have three
($\degree{60}$, $\degree{180}$ and $\degree{300}$).

For each carbonated oscillator the correlations with the
hydrocarbonat oscillators coupled to it will be studied.  For example, a
correlation will be searched between a CC bond and the four coupled CH
bonds.  This will determine whether or not the energy of the CH bonds
should participate in the energy of the GG grain corresponding to a CC
bond.

Two cases can be distinguished, depending on whether the potentials
considered are harmonic or have a triple-cosine form.

\subsection {Harmonic Potentials\label{determination-pot-harmonique}}

We consider the case of two coupled harmonic oscillators, one
hydrocarbon of the $\alpha$ type (e.g., CH) and the other carboned
of type $\beta$ (for example, CC). The correlation between the two is
represented by a {\it correlation factor} $\corr X$, with a value
between $-1$ and $1$, linking their average values $m_\alpha$ and
$m_\beta$ and verifying:
\begin{equation}
  (m_\alpha -\equi \alpha) / \equi \alpha
  =
  \corr X (m_\beta - \equi  \beta ) / \equi \beta
  \label {facteur-correlation}
\end{equation}
or equivalently:
\begin{equation}
  m_\alpha -\equi \alpha
  =
  \corr X (\equi \alpha / \equi \beta) (m_\beta - \equi  \beta ) 
\end{equation}
If $\corr{X} = \pm 1$, then one has $m_\alpha = \pm (\equi {\alpha} / \equi
{\beta}) m_\beta$. In this case, there is a total correlation between $m_\alpha$ and
$m_\beta$.
If $\corr{X} = 0$ then $m_\alpha$ is in no way dependent on $m_\beta$
because $m_\alpha = \equi {\alpha}$.
In this case, we have a
total non-correlation.  In the other cases, the correlation
is only partial.

Let us recall that the UA grain centers correspond to the AA carbon
atoms and that the GG and CC equilibrium bond lengths are equal
($\equi {CC} = \equi {GG}$).

\paragraph{Bond.}

Let $\corr {A}$ be the correlation factor of the CH and CC bonds.
We define the {\it average energy contribution} $\cal C_{CH}$ of a coupled CH bond
to the energy of a GG bond of length $l$ by:
\begin{equation}
{\cal C}_{CH} (l) = k_{CH} . \corr {A}^2 . (\equi {CH} / \equi
{CC})^2 . (l - \equi {CC})^2
\end{equation}
The potential energy of a GG bond of length $l$ is then the potential energy of the
corresponding CC bond, plus the average contributions of the 4 CH
bonds that are coupled to it, for the same length $l$~:
\begin{equation}
{\cal U}_{GG} (l) = {\cal U}_{CC} (l) + 4. {\cal  C}_{CH} (l)
\end{equation}
This implies:
\begin{equation}
  {\cal U}_{GG} (l) = k_{CC} . (l - \equi {CC})^2
                             + 4. k_{CH} .  \corr {A}^2 .  (\equi {CH} / \equi {CC})^2 .  (l - \equi {CC})^2  
\end{equation}
 whence~: 
\begin{equation}
  {\cal U}_{GG} (l) = (
               k_{CC} + 4 . k_{CH} . \corr {A}^2 .
             (\equi {CH} / \equi {CC})^2 )
             .
             (l - \equi {CC})^2
\end{equation}
which means (since $\equi {CC} = \equi {GG}$) that the stiffness
constant of the bond GG is:
\begin{equation}
  k_{GG} = k_{CC}
              + 4 . k_{CH} . \corr {A}^2 . (\equi {CH} / \equi  {CC})^2                \label {eq:kcc}
\end{equation}

\paragraph{Valence Angle.}
The treatment of valence angles is very similar in that, like bonds,
their potential is harmonic. A difference is however due to the
existence of two types of oscillators of valence angles involving
hydrogens (CCH and HCH) while there exists only one for bonds
(CH). Furthermore, correlations with a CCC angles can be different for
CCH angles and for HCH angles. Let $\corr {B}$ be the correlation
factor between the angle CCH and the angle CCC, and let $\corr {C}$ be
the correlation factor between the angle HCH and the angle CCC. Four
CCH bonds and only one HCH bond are coupled to CCC.  The stiffness
constant of the GGG potential is thus:
\begin{equation}
  k_{GGG} = k_{CCC}
           + 4 . \corr {B}^2 . ( \equi {CCH} / \equi {CCC} )^2 . k_{CCH}
           +      \corr {C}^2 . ( \equi {HCH} / \equi {CCC} )^2 . k_{HCH}
  \label {eq:kccc}
\end{equation}

\subsection{Non-Harmonic Potentials}

The potentials of the dihedral angles do not have a single equilibrium
value, as is the case for harmonic potentials, but
three minima distant from $\degree {120}$.  So we can return to the case of a
single minimum, considering the values modulo
$\degree {120}$ of the dihedral angles.
Applying the same technique as for the valence angles to
each of the 3 parameters of the equation (\ref {equation:triple-cosine}),
we get, for $i = $1,2,3:
\begin{equation}
  \paramcos{GGGG} {i} = \paramcos{CCCC} {i}
                    + 4 . \corr {D}^2 . \paramcos {CCCH}{i}
                    + 4 . \corr {E}^2 .  \paramcos {HCCH}{i}  
 \end{equation}
 The quotients of the equilibrium values have disappeared: they are
 all equal to 1 since the equilibrium value is the same for all the
 dihedral angles modulo $\degree {120}$. $\corr {D}$ is the
 correlation factor between the 4 CCCH angles coupled to the CCCC
 angle, and $\corr {E}$ is the correlation factor between the four
 angles HCCH coupled with the CCCC angle.  The potential
 ${\cal U}_{GGGG}$ is then written as:
\begin{equation}
  {\cal U}_{GGGG} (\phi) =
      {\cal U}_{CCCC} (\phi)
  + 4.\corr {D}^2. {\cal U}_{CCCH} (\phi)
  + 4.\corr {E}^2. {\cal U}_{HCCH} (\phi)
\end{equation}

\section {Determination of UA potential\label {section:determination-potentiel-UA}}
Several simulations in molecular dynamics have been carried out with
various molecules, ranging from $\alkane {6}{12}$ to 
$\alkane {24}{48}$. In every simulation, the molecule in
question is isolated in the vacuum. Thus, for a given temperature, its
dynamics depends only on the intra-molecular potentials. The
simulations were all ran with the dynamics molecular software
described in \cite{RPSP-IJMPC}.

The main simulation is run at $\temp {300}$, a temperature which
allows CCCC dihedral angles to reach the conformations {\gauche} and
{\gauchep} from an initial  {\it all-trans} conformation. 
Additional simulations have also been carried out between
$\temp {10}$ and $\temp {1000}$ to explore the behaviour of the
molecules, on the one hand close to equilibrium, in conformation
{\trans} ($\temp {10}$), and on the other hand by getting closer of
the free-rotation condition ($\temp {1000}$).

One always wait for the molecule to reach a state of dynamic
equilibrium to begin to study the correlations between the oscillator
kinematics.  This state of dynamic equilibrium implies that the
average temperature of each of the atoms no longer evolves in the
course of time and therefore that the same applies to the average
kinetic energy of bonds, valence angles and dihedral angles.  It is
then possible to study the energy exchanges between these oscillators,
exchanges that are responsible for the correlations between their kinematics.

The initial energy of the molecule is provided by stretching the first
CC bond, while the rest of the molecule is in the CC equilibrium state.
The potential energy thus supplied
propagates to all atoms and gives the average temperature of the
molecule. The transition period before reaching 
dynamic equilibrium can be quite long and for example be $\ns {1}$, or
$10^6$ time steps, for an $\alkane {24}{48}$  molecule.  By default,
it is this transition period of $\ns 1$ that is imposed on all the
simulations.

Once dynamic equilibrium is reached, the state of the molecule
(positions of the atoms, lengths of the bonds, values of the valence
and dihedral angles) is recorded every femto-seconds, for 50,000
time-steps, or $\ps {50}$. This is from these recorded values that the
correlation analysis is conducted.

\subsection {Correlation Analysis of CC Bonds}
The first step has been to determine whether there is a correlation between the bonds
$\CC {2}{3}$ and $\CH {3}{13}$ of the molecule $\alkane{6}{12}$.
The resulting cloud of points has an elliptical shape (\myfig
{\ref{correl-bond}}).
\begin{myfigure}
\begin{center}
\includegraphics [width=7.5cm] {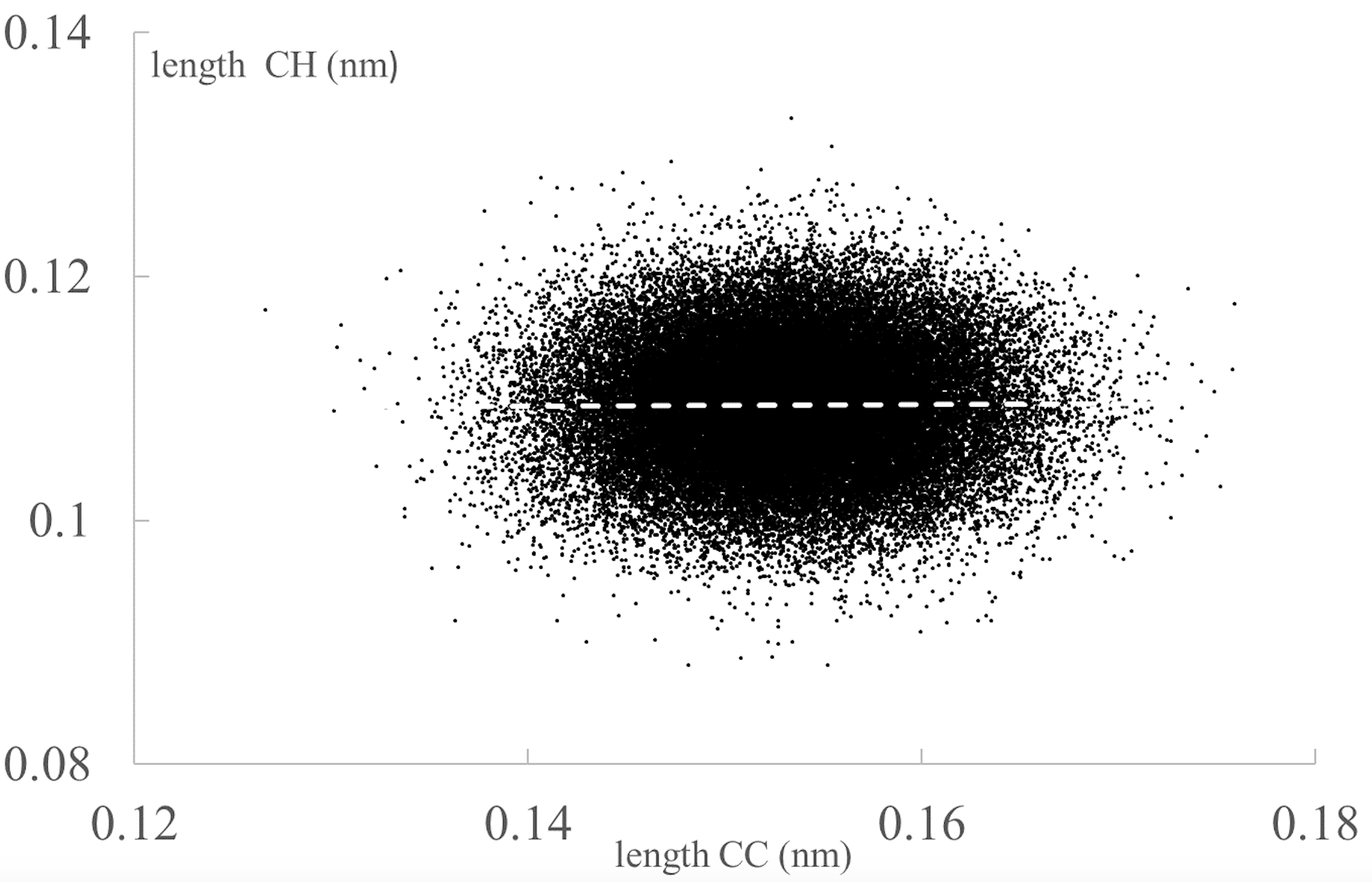}
\caption{\small Correlation of length $\CH {3}{13}$ according to
  the length $\CC {2}{3}$. Dashed regression line.
}
\label{correl-bond}
\end{center}
\end{myfigure}
The distribution of points is apparently random within the cloud. The
elliptical shape is due to the difference in equilibrium lengths of CC
bonds ($\equi{CC} = \nm {0.1529}$) and CH bonds
($\equi{CH} = \nm {0.109}$).  The magnitude of the distribution is
directly related to the temperature and therefore to the amplitude of
variation of the bond lengths relative to their equilibrium value.
The treatment of this distribution by a linear regression of least
squares method gives a zero slope, with a precision of $10^{-5}$.  The
same analysis has been applied to the 3 other CH bonds coupled to
$\CC {2}{3}$ ($\CH {2}{10}$, $\CH {2}{11}$, $\CH {3}{12}$) with an
identical result. More generally, there is an absence of correlation
between $\CC {2}{3}$ and any CH bond of the molecule.  The lack of
correlation with the coupled CH bonds has also been checked for the
other CC bonds ($\CC {0}{1}$, $\CC {1}{2}$, $\CC {3}{4}$ and
$\CC {4}{5}$).  Thus, the result is always the same~: there is no
correlation in time between the lengths of the CC and CH bonds.  This
results have been confirmed on longer molecules.  Therefore, the
correlation factor $\corr{A}$ in equation (\ref{eq:kcc}) is equal
to 0 which implies that the CC and GG bond potentials are identical:
\begin{equation}
k_{GG} = k_{CC} = \kcalMolAngsquare {268} \label {eq:param-gg}
\end{equation}

\subsection{Correlation Analysis of CCC Valence Angles}

The same method is applied to the correlation analysis of the angle of
valence $\CCC {2}{3}{4}$ of the molecule $\alkane{6}{12}$ with the
CCH and HCH angles.  It can be seen that the values of the angles that
are uncoupled to $\CCC {2}{3}{4}$ have a zero correlation with this one.
On the other hand, angle $\HCH {12}{3}{13}$ (the HCH angle attached to carbon $C_3$)
has a non-zero correlation with the angle $\CCC {2}{3}{4}$.  The
corresponding cloud of points is shown on the left side of 
\myfig {\ref{correl-val}}.

The regression slope is $- 0.21 \pm 0.003$.  The negative value of
the slope means that, on average, the HCH angle has a value less than its equilibrium value
$\equi {HCH}$ when the CCC angle has a value greater than
its equilibrium value $\equi {CCC}$.  The value of the slope of
regression equals $\corr{C}.\equi {HCH} / \equi {CCC}$.  Like the
ratio of equilibrium angle values $\equi {HCH} / \equi {CCC}$
is $108.4 / 112.7 = 0.962$, the correlation coefficient
$\corr {C}$ is $-0.21 / 0.962 = -0.218$.
As a result, the
correlation is only partial, which implies that the energy of the
the angle $\HCH {12}{3}{13}$ only partly participates in the energy
of the angle $\CCC {2}{3}{4}$.  Of all the CCH angles, only the
four angles $\CCH {2}{3}{13}$, $\CCH {2}{3}{14}$, $\CCH {3}{4}{13}$
and $\HCC {14}{3}{4}$ have a non-zero correlation with the valence angle
$\CCC {2}{3}{4}$ (the case of $\CCH {2}{3}{13}$ is shown
in the right side of \myfig {\ref{correl-val}}).  We can see
that the regression slope is the same for all four CCH angles and
is $-0.2 \pm 0.003$.  The regression slope is equal to
$\corr{B}.\equi {CCH}/\equi {CCC}$.  Like the ratio
$\equi {CCH} / \equi {CCC}$ is $110.5/112.7 = 0.98$, the correlation factor
$\corr {B}$ is $-0.2 / 0.98 = -0.204$. 
Therefore, this correlation is also partial.  The equation
(\ref {eq:kccc}) is then written:
\begin{equation}  
k_{GGG} = 58.35 + 4(-0.204)^2 \times (108.4/112.7)^2 \times 37.5 +
(-0.218)^2 \times (110.5/112.7)^2  \times 33
\end{equation}
thus:
\begin{equation}  
k_{GGG} = \kcalMolAngsquare {65.63} \label {eq:param-ggg}
\end{equation}

\begin{myfigure}
\begin{center}
  \includegraphics [width=7.5cm] {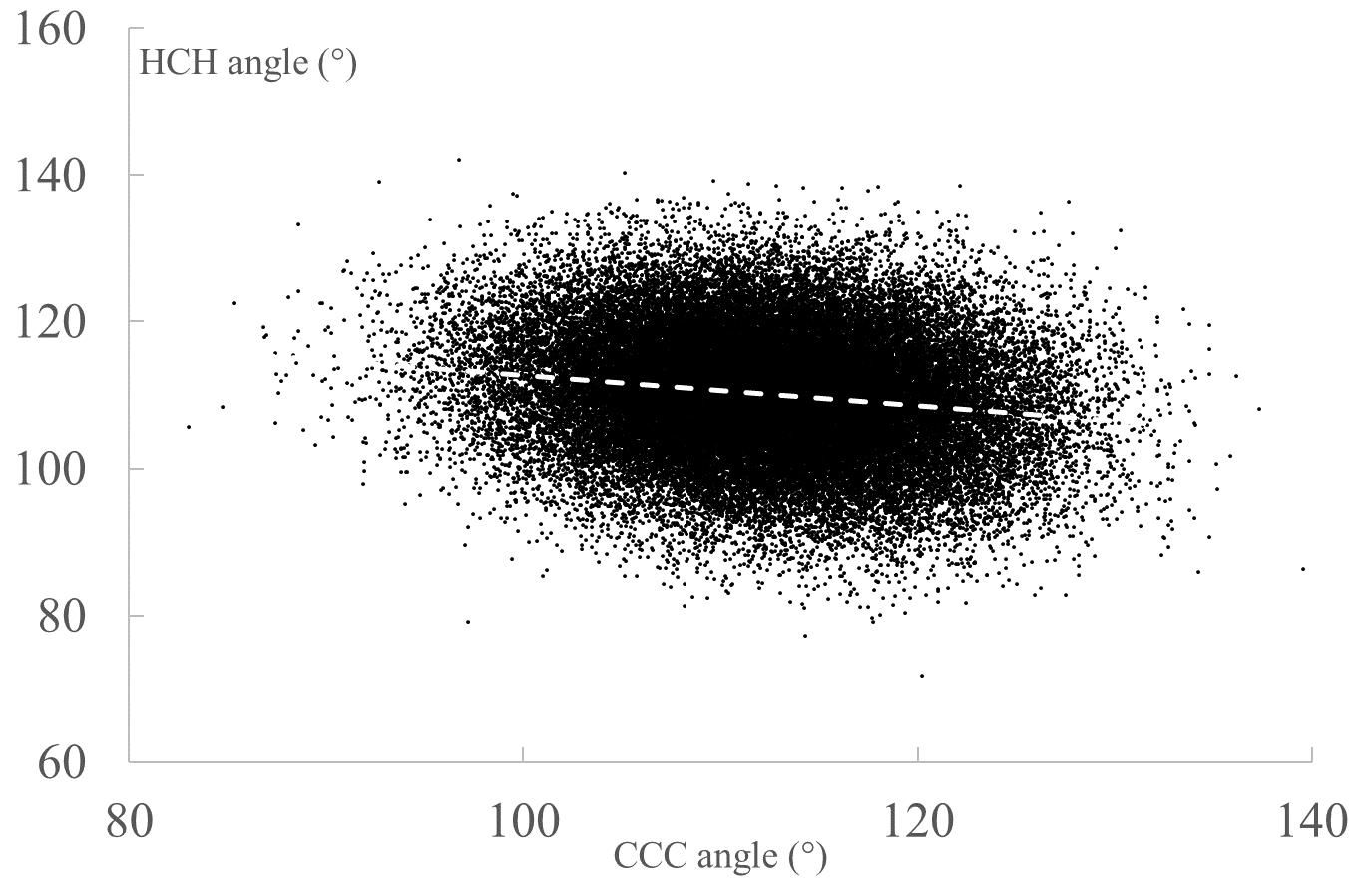} ~~~
  \includegraphics [width=7.5cm] {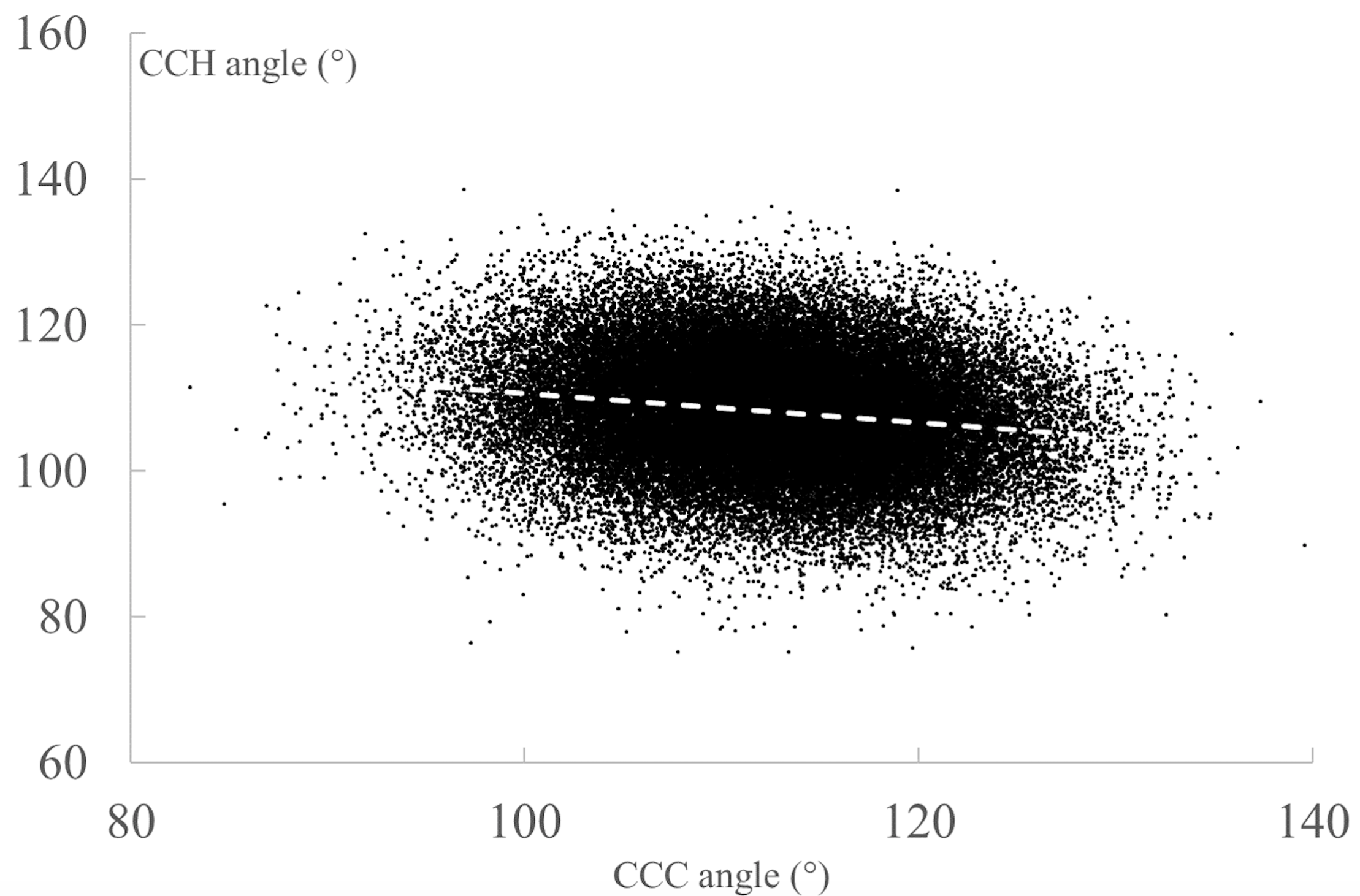}
  \caption{\small {\bf Left Image:} correlation of valence angle $\HCH {12}{3}{13}$
  with the angle $\CCC {2}{3}{4}$. Dashed regression line.
  {\bf Right Image:}
correlation of valence angle $\HCH {2}{3}{13}$
  with the angle $\CCC {2}{3}{4}$.
}
\label{correl-val}
\end{center}
\end{myfigure}

\subsection{Correlation Analysis of CCCC Dihedral Angles}
We now apply the correlation analysis between a dihedral angle
CCCC and the dihedral angles comprising at least one hydrogen, HCCH and
CCCH.  As a reference, we retain the angle $\CCCC {1}{2}{3}{4}$ that
is coupled to the following angles:
$\HCCC {10} {2}{3} {4}$,
$\HCCC {11} {2}{3} {4}$, $\CCCH {1} {2}{3} {12}$,
$\CCCH {1} {2}{3} {13}$, $\HCCH {10} {2}{3} {12}$,
$\HCCH {11} {2}{3} {12}$, $\HCCH {10} {2}{3} {13}$ et
$\HCCH {11} {2}{3} {13}$.

\begin{myfigure}
\begin{center}
  \includegraphics [width=7.5cm] {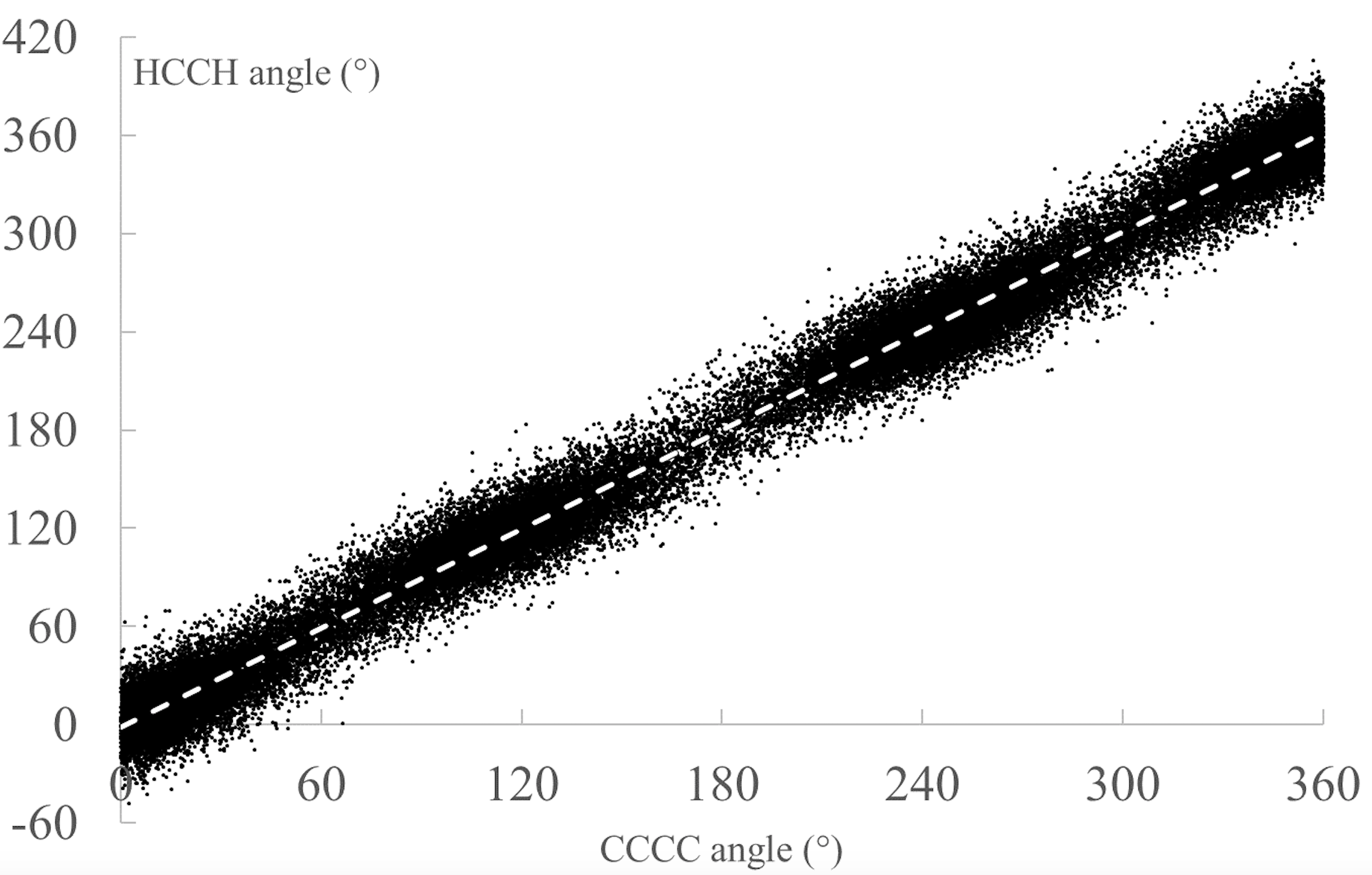} ~~~
\includegraphics [width=7.5cm] {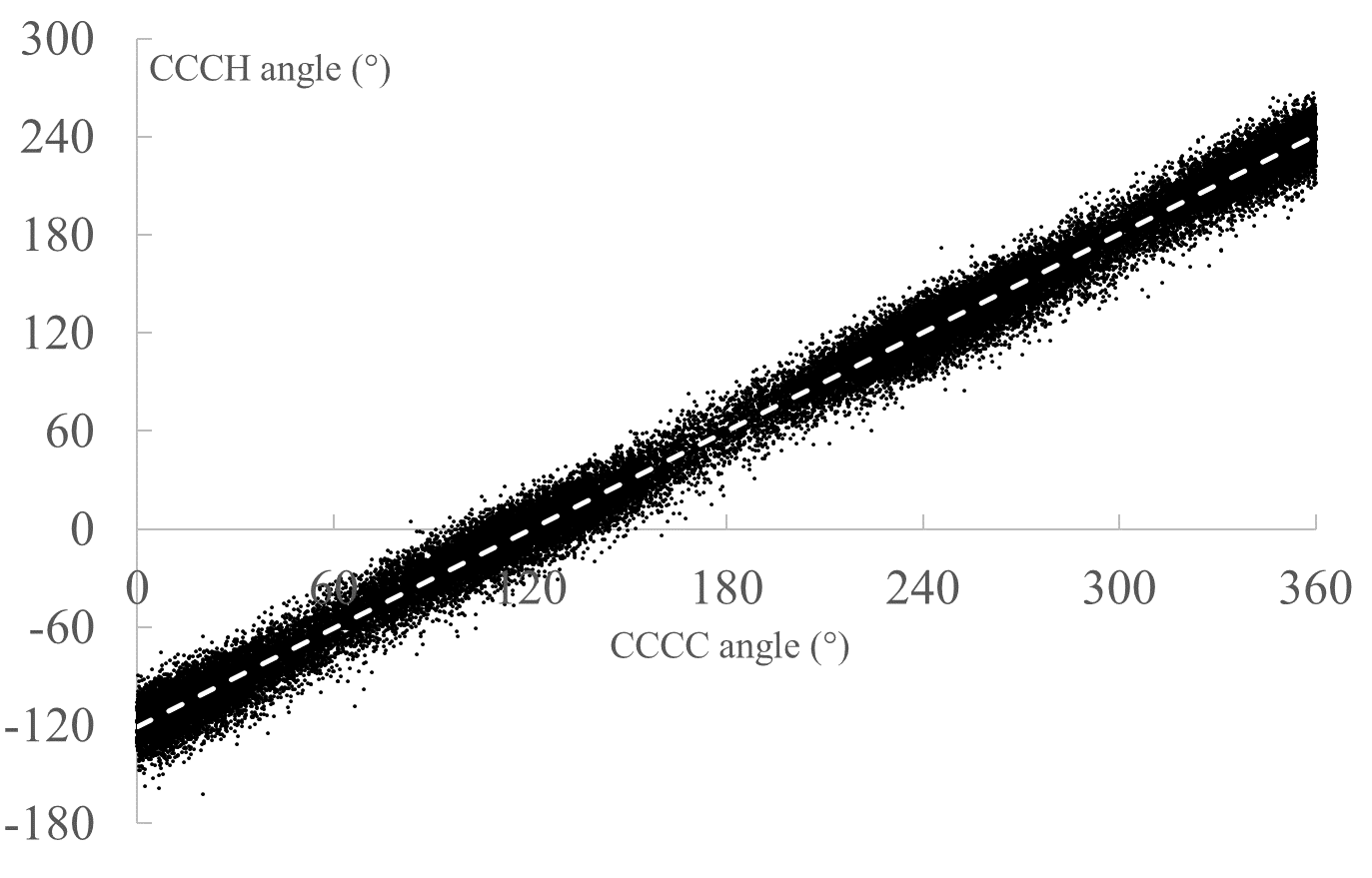}  
\caption{\small {\bf Left Image:}
correlation of $\HCCC {11}{2}{3}{4}$
with the dihedral angle $\CCCC {1}{2}{3}{4}$ (dashed regression line).
{\bf Right Image:}
correlation of $\HCCH {11}{2}{3}{13}$
with the dihedral angle $\CCCC {1}{2}{3}{4}$.
}
\label{correl-dih}
\end{center}
\end{myfigure}
The first striking fact is that all the analyses give a slope
regression line equal to $1 \pm 0.007$.  So there is an equal
relationship between the angular variations of the angle
$\CCCC {1}{2}{3}{4}$ and the variations of the coupled angles CCCH and HCCH.  The
distribution of the angular values on either side of the regression
line is related to the kinetic energy of vibration exchanged with the
other dihedral angles.

The second striking fact is that the angles considered only differ by
their angular phase-shift, given by their ordinate at origin.

The
angular shifts, resulting from linear regression, have a
periodicity of $\degree {120}$ (Table \ref{dephasage}).  The angles
are perfectly correlated: when the CCCC angle varies, it 
directly drives the eight dihedral angles coupled to it. One thus has:
$\corr {D} = \corr {E} = 1$.
\begin{table}[ht]
\begin{center}
\begin{tabular}{c|ll}
angular phase-shift & dihedral angles \\
\hline
$-\degree {120}$ &  $H_{11}C_{2}C_{3}C_{4}$  ~~~~ $C_{1}C_{2}C_{3}H_{14}$
           ~~~~ $H_{12}C_{2}C_{3}H_{13}$\\
$\degree {0}$      &   $H_{11}C_{2}C_{3}H_{13}$   ~~ $H_{12}C_{2}C_{3}H_{14}$\\
$\degree {120}$  &  $H_{12}C_{2}C_{3}C_{4}$   ~~~~   $C_{1}C_{2}C_{3}H_{13}$   ~~~~  $H_{11}C_{2}C_{3}H_{14}$
\end{tabular}
\caption{\small Phase-shifts of the dihedral angles coupled to
  $C_{1}C_{2}C_{3}C_{4}$}.
\label {dephasage}
\end{center}
\label{default}
\end{table}

The perfect correlation between a CCCC dihedral angle and the dihedral
angles coupled to it is also checked for the two other CCCC dihedral
angles of the $\alkane{6}{12}$ molecule.  One
also observes the same perfect correlation in the case of
longer molecules, which makes the result quite
general. Because of this correlation and of the modulo
$\degree {120}$, we got:
\begin{equation}  
{\cal U}_{GGGG} (\phi) =  {\cal U}_{CCCC} (\phi) + 4 . {\cal
  U}_{CCCH} (\phi) + 4 . {\cal U}_{HCCH} (\phi)
\end{equation}  
Hence finally:
\begin{eqnarray}
  & \paramcos{GGGG}{1} = \paramcos{CCCC}{1} = 1.74~kcal / mol     \nonumber \\
  & \paramcos{GGGG}{2} = \paramcos{CCCC}{2} = - 0.157~kcal / mol \label {eq:param-gggg} \\
  & \paramcos{GGGG}{3} = \paramcos{CCCC}{3} +4 . \paramcos{CCCH}{3} +4 . \paramcos{HCCH}{3}
  = 3.015 ~ kcal / mol \nonumber
\end{eqnarray}

\paragraph {Conclusion.}
The three components of the intra-molecular UA potential (GG bonds,
GGG valence angle, and GGGG dihedral angle) were determined (Eq. (\ref
{eq:param-gg}), (\ref {eq:param-ggg}), (\ref {eq:param-gggg}) by a
correlation analysis.

A CC bond oscillator is totally decorrelated in time of the CH
oscillators.  This implies that the vibration of the GG oscillator is
the same as that of the CC oscillator.  Thus, the potential energies
of the CC and of the GG oscillators are identical.  There is no contribution
of the potential energy of the CH oscillators to the potential energy
of the GG bond.

On the contrary, the dihedral angle oscillators CCCC and CCCH as well
as HCCH are totally correlated.  Changing the value of a CCCC angle
results in the same angular variation of the CCCH and HCCH that are
coupled with it.  Therefore, the potential energy of a dihedral angle
GGGG is the sum of the potential energies of the corresponding CCCC
angle with the potential energies of the dihedral angles CCCH and HCCH
coupled to it (sharing with it the two central carbons).

Valence angles present an intermediate case.  Because of the weak
coupling ($\corr {B} \approx \corr {C} \approx -0.2$), the GGG and CCC
potentials are different, but only a small percentage of the
contribution to GGG's potential energy is made by the CCH and HCH
angles.  It should be noted that this is not a driving effect since,
in fact, the CCC angles and the CCH and HCH angles vary in the opposite
direction.

It should also be noted at this point that if the masses driven by the grains during the
UA simulations were depending on the previous correlations, these masses
should be different for bonds and for valence angles.  That
would certainly be a contradiction in terms on a physical level.
This issue will be considered in more detail in the next section.

\section {Validation of the UA Potential\label{validation}}
The main objective of UA simulations is to obtain kinematics
in UA as close as possible to those in AA.
This proximity of kinematics is of course only sought for the same
chain length, at the same temperature, and starting from the same initial condition.
Thus, we will compare the kinematics of an AA molecule and that
of the corresponding UA molecule, at the same temperature, the UA grains
being initially at the same positions as the carbon atoms of the AA molecule.

One immediately notes that, because of the number of atoms involved, the global kinematics
of the molecules at both scales cannot be strictly identical, except for very short periods
of time (consequence of Poincaré's three-body theory \cite{trois-corps}; we are in a framework
of {\it deterministic chaos}).  In fact, the divergence of the kinematics results
from a minute difference, either in the initial conditions, or in the parameters
of the UA potentials.  A very small difference of one of these values
(> $10^{-10}$) causes a divergence in a delay of the order of $\ps {10}$.

Only average values from the kinematics can therefore be compared.
Two kinematics, one in AA and the other in UA, will be consistent if they explore
the same phase space.
From this point of view, several comparison criteria are possible:
\begin{enumerate}

\item
  Comparison of the average periods of vibration of the corresponding harmonic
  oscillators (e.g. CC bond and corresponding GG bond). 

\item
  Comparison of mean energies obtained by a Boltzmann-inverse approach
  using probabilities of presence as a function of length (bond) or angle
  (valence or dihedral), at both scales.

\item
  Comparison of the mean end-to-end distances (distances between
  the two carbons or grains at the extremities) of an AA molecule
  and of the corresponding UA molecule.

\end{enumerate}
It is important to note that these three criteria are independent of each
other and that the first two are at the local scale (level of each oscillator),
while the third is at the global scale (complete molecule).

\subsection {Comparison of Mean Vibration Periods\label{vibration_period}}
The period of vibration $T$ of a harmonic {\it isolated} oscillator composed of two mass objects $m_1$ and $m_2$ is defined by:
\begin{equation}
T = 2\pi \sqrt{0.5 \times m_r / k} \label{period-harmonic}
\end{equation}
where $m_r$ is the reduced mass $m_r = 1 /(1/m_1 + 1/m_2)$ and $k$ is the stiffness constant of the oscillator. In the case of an oscillator composed of two carbons, we have $m_r = \gmol {6}$ and
  $k = \kgMolNmsquare {112.1312}$ $( \kcalMolAngsquare {268} )$ \cite{OPLS}.
The period of this isolated CC oscillator is thus $T = \fs {32.45}$, and this whatever the temperature.

Using a Fourier transform ({\FFT}), the period of vibration of the isolated CC oscillator can be extracted and measured from simulation data. Thus, from a simulation (at $\temp {300}$) generating $2^{16}$ measurements (time-steps), the measured vibration period, corresponding to the single vibration peak, is indeed the expected one of $\fs {32.45}$ (\myfig {\ref{CC-FFT}}).
\begin{myfigure}
  \begin{center}
    \includegraphics [width=8cm] {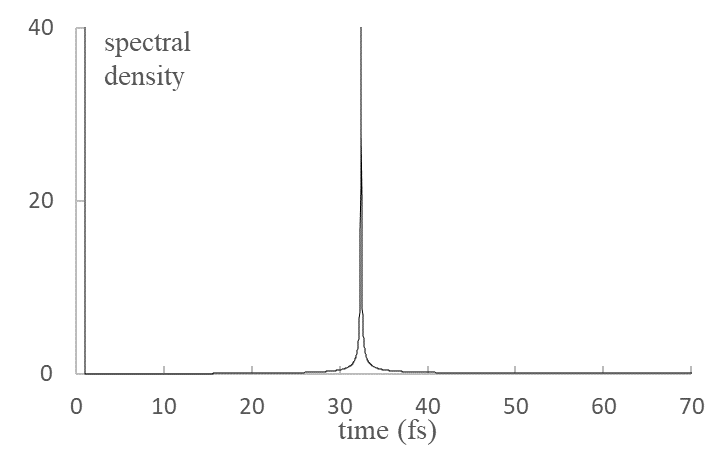} 
\caption {\small Analysis by {\FFT} of the isolated CC oscillator.
}
\label{CC-FFT}
\end{center}
\end{myfigure}

It's a different situation with a harmonic {\it non-isolated} oscillator. In this case, indeed, several peaks appear in the Fourier transform and we can no longer speak of "the" period of vibration of the oscillator.
The simplest case of non-isolated oscillator is the CC bond within a $\alkane {2}{4}$ molecule. An analysis by {\FFT} of the molecular dynamics of this molecule at $\temp {20}$ shows 3 peaks (left image of the \myfig {\ref{C2H4-20}}).
\begin{myfigure}
  \begin{center}
    \includegraphics [width=8cm] {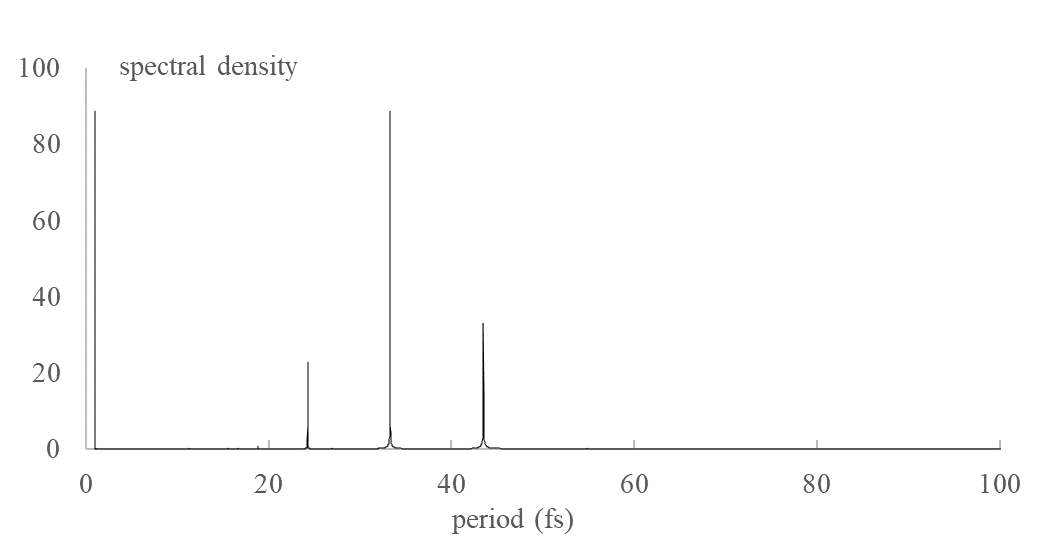} 
    \includegraphics [width=8cm] {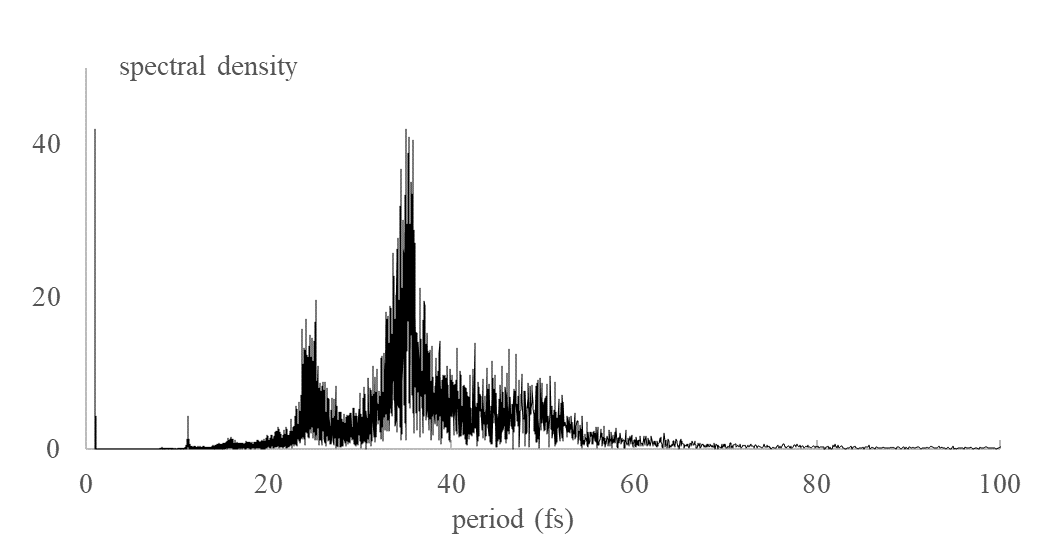}    
    \caption{\small Analysis by {\FFT} of the dynamics of the CC bond in an $\alkane {2}{4}$ molecule.
      {\bf Left Image:} temperature is $\temp {20}$. {\bf Right Image:} temperature is $\temp {1000}$.
}
\label{C2H4-20}
\end{center}
\end{myfigure}
The main peak corresponds to a period of $\fs {33.3}$, i.e. a value significantly different from that of the isolated CC oscillator ($\fs {32.45}$).
The number of peaks increases with the temperature (the right image of the \myfig {\ref{C2H4-20}} shows the peaks obtained at $\temp {1000}$).

The number of peaks also increases with the size of the molecule.  Various molecules from $C_{2}H_{4}$ to $C_{100}H_{200}$, various CC bonds in these molecules and various temperatures were considered. The \myfig {\ref{C100H200-100}} shows the peaks obtained with the molecule $C_{100}H_{200}$ at both temperatures of $\temp {100}$ and $\temp {1000}$.
\begin{myfigure}
  \begin{center}
    \includegraphics [width=8cm] {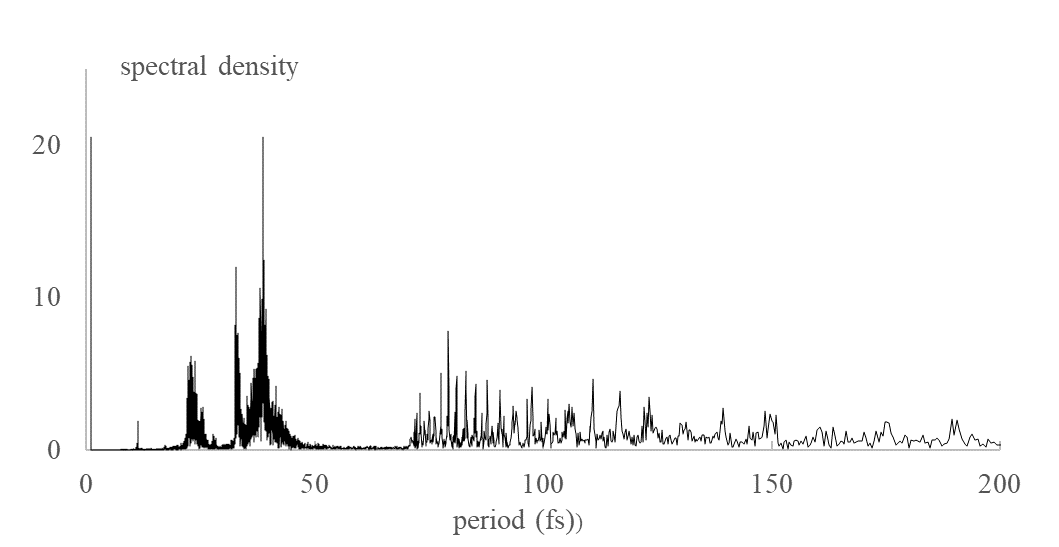} 
    \includegraphics [width=8cm] {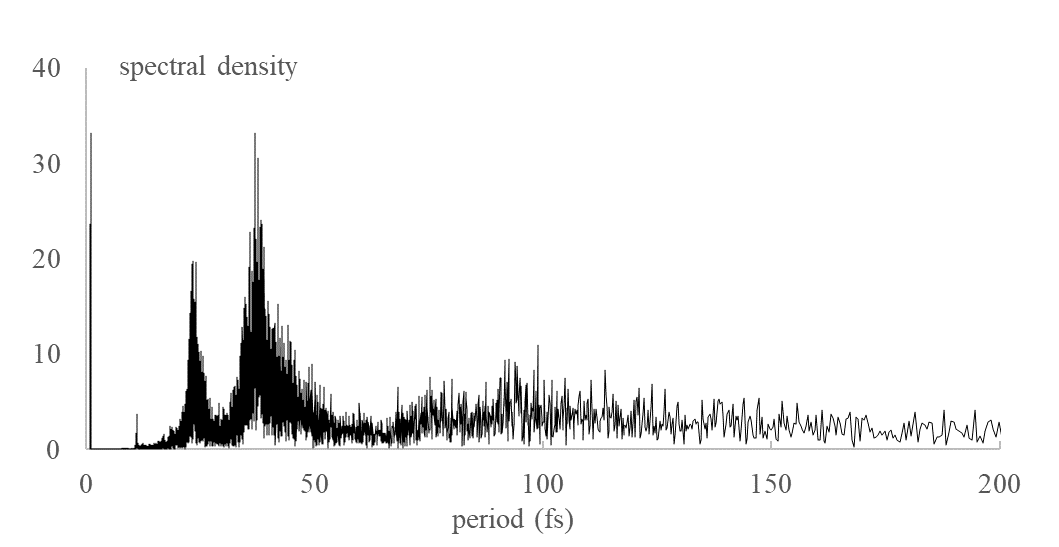}        
    \caption{\small
Analysis by {\FFT} of the dynamics of the bond $C_1C_2$ in a molecule $\alkane {100}{200}$.
      {\bf Left Image:} temperature is $\temp {100}$. {\bf Right Image:} temperature is $\temp {1000}$.
}
\label{C100H200-100}
\end{center}
\end{myfigure}

If we cannot speak of "the" period of a non-isolated harmonic oscillator, we are nevertheless going to define its {\it mean period} which is that of an isolated harmonic oscillator {\it equivalent}, in a sense that we are going to specify now.

The starting point is the Fourier transform of a non-isolated harmonic oscillator whose vibration frequency is considered to be disturbed by interactions with other oscillators.  These interactions lead to a degeneration of the vibration peak of the isolated oscillator producing a multiplicity of $N$ peaks.  Thus, each peak corresponds to an "interaction mode" of the oscillator with the other oscillators and its spectral density $D_i$ reflects the probability $P_i$ that the oscillator is in the "mode" corresponding to the peak. The probabilities are defined by:
\begin{equation}
  P_i = D_i / \sum_1^N {D_j}
\end{equation}
Each natural frequency $T_i$ corresponds to a sinusoidal signal in time which can be considered as produced by a fictitious isolated harmonic oscillator. It has the same stiffness $k$ as that of the non-isolated oscillator, and whose mass $m_i$, called {\it fictitious reduced mass}, is obtained from the equation (\ref{period-harmonic}) by:
  \begin{equation}
  m_i = kT_i^2 / 2\pi^2
\end{equation}

The {\it dynamic reduced mass} $mr_D$ of a non-isolated oscillator is then defined as the sum of the fictitious reduced masses of the various fictitious oscillators, weighted by their probability $P_i$:
\begin{equation}
 mr_{D} = \sum_1^N {P_i  m_i} = (k/2\pi^2) \sum_1^N {P_i T_i^2}
\end{equation}
The {\it dynamic mass} $m_D$ is twice the dynamic reduced mass:
\begin{equation}
  m_D = 2 \times mr_D
\end{equation}
The average period $T_D$ of a non-isolated harmonic oscillator is then defined as that of the isolated harmonic oscillator of the same stiffness and whose mass is the dynamic reduced mass of the non-isolated oscillator:
\begin{equation}
  T_D = 2\pi \sqrt {0.5 \times mr_D / k} \label{period-mean}
\end{equation} 

\paragraph {Results.} 
All the {\FFT} analyses performed give for the AA bonds dynamic masses having the same value of $14 \pm 0.2~g / mol$ with an average period $T = \fs {35.05}$, whatever the temperature.

The determination of dynamic masses can also be carried out at the UA level. For example, we find that the dynamic mass of a GG bond is in all cases (different molecule lengths and different temperatures) $\gmol {14}$. Thus, the dynamic masses of the bonds are the same in AA and UA. With equal stiffnesses in UA and AA, the average periods of AA and GG bonds are the same. This result means that, with equal stiffnesses in AA and UA, the first (local) criterion for comparison between AA and UA is verified for bonds.

A similar analysis of the kinematics of the CCC valence angles in AA also leads to the conclusion that the average dynamic mass of the valence angles is $\gmol {14}$. With approximately the same stiffnesses of the CCC and GGG valence angles, the mean periods of the valence angles are very close in AA and UA, so that the first criterion for comparison between AA and UA is also verified for the valence angles.

If, instead of using the dynamic mass of the oscillators, we took the static mass ($\gmol{12}$ in AA and $\gmol{14}$ in UA), we could think that, to obtain equal periods, we would simply fix the stiffness $k'_{GG}$ of the GG bond as $k'_{GG} = (14/12) \times k_{CC} = \kcalMolAngsquare {312.67}$.
This would give us a new UA potential, called {\it modified potential}. Several simulations were carried out using this modified potential and the average vibration period obtained is $\fs {32.5}$ which does not correspond to the average period observed during the simulations of aliphatic molecules in AA. The modified potential is thus not the desired potential because it makes the mean periods of the CC and UA bonds different, leading to divergent kinematics.

In conclusion, the fact that in AA and UA, on the one hand, the stiffness of the bonds are equal and, on the other hand, that the stiffness of the valence angles are equal has the effect of making the average periods of the corresponding oscillators equal. This is the first (local) criterion for comparison between AA and UA.


\subsection{Comparison of Energies Obtained by Boltzmann-Inverse}
Oscillators, whether AA or UA, are now analyzed by a Boltzmann-inverse method.  In this approach, the energy of an oscillator is obtained from the densities of probability of presence in a given state.
The formula defining the energy $U$ is~:
\begin{equation}
U = k_B \times T \times ln (P_0 / P_i)
\end{equation}
where $k_B$ is the Boltzmann constant, $T$ is the temperature of simulation, $P_0$ is the maximum probability and $P_i$ is the probability of state $i$.  The treatment of bonds and valence and dihedral angles is performed from the same simulation in molecular dynamics. The range of variation of the values has been segmented into 200 evenly distributed classes and the probabilities of presence in each class have been obtained from $10^6$ time-steps.

The energies of the CC and GG bonds, of the CCC and GGG valence angles, and of the CCCC and GGGG dihedral angles, were compared for the two molecules $\alkane{16}{32}$ and $G_{16}$, at a temperature of $\temp{300}$. 
The same analysis was also performed, with the modified potential defined in the section \ref{vibration_period}.
The results obtained are shown in \myfig{\ref {boltzmann}}.
\begin{myfigure}
\begin{center}
  \includegraphics [width=7cm] {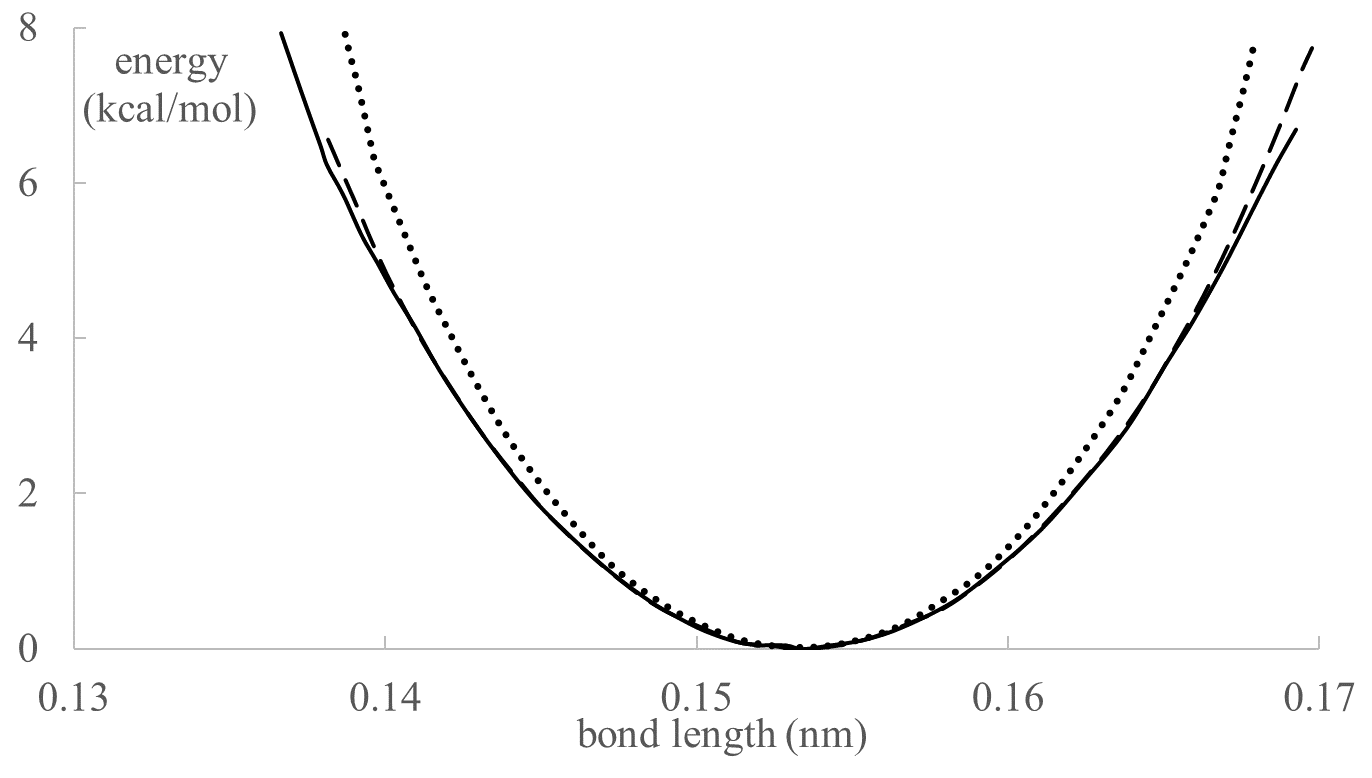}      ~~~~~
  \includegraphics [width=7cm] {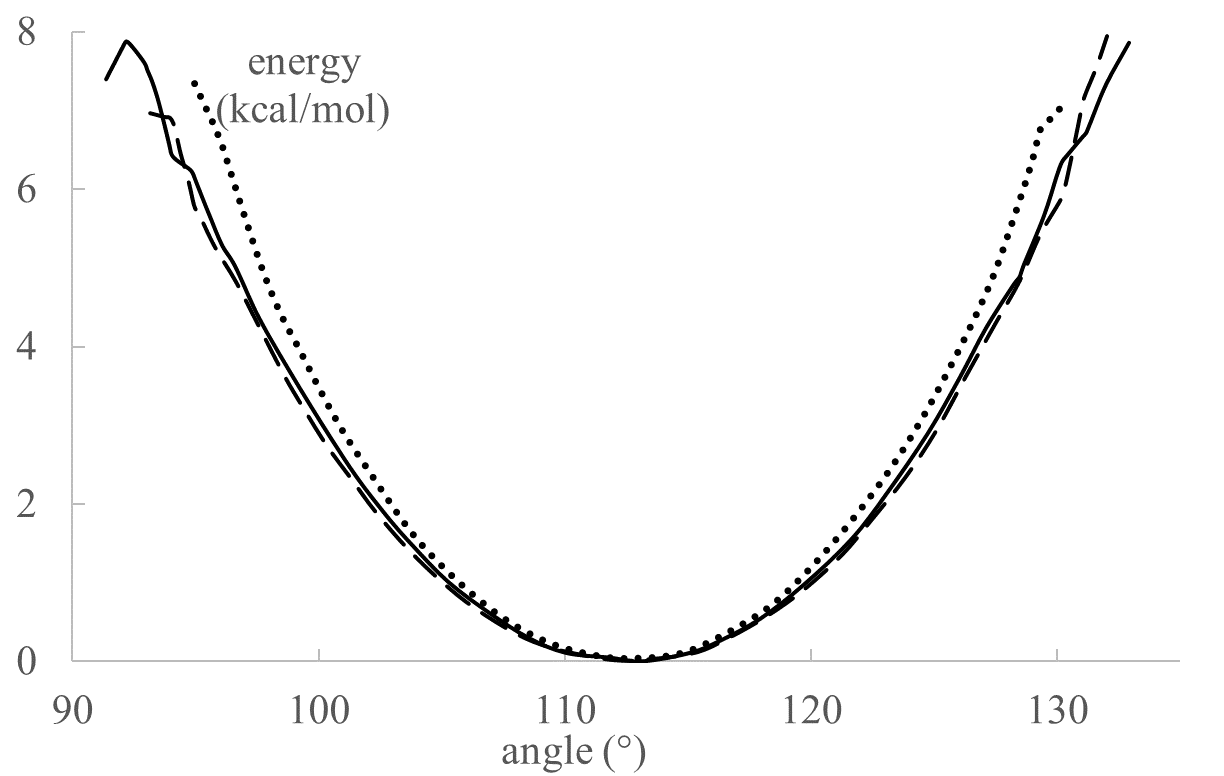} ~~~~~
\includegraphics [width=7cm] {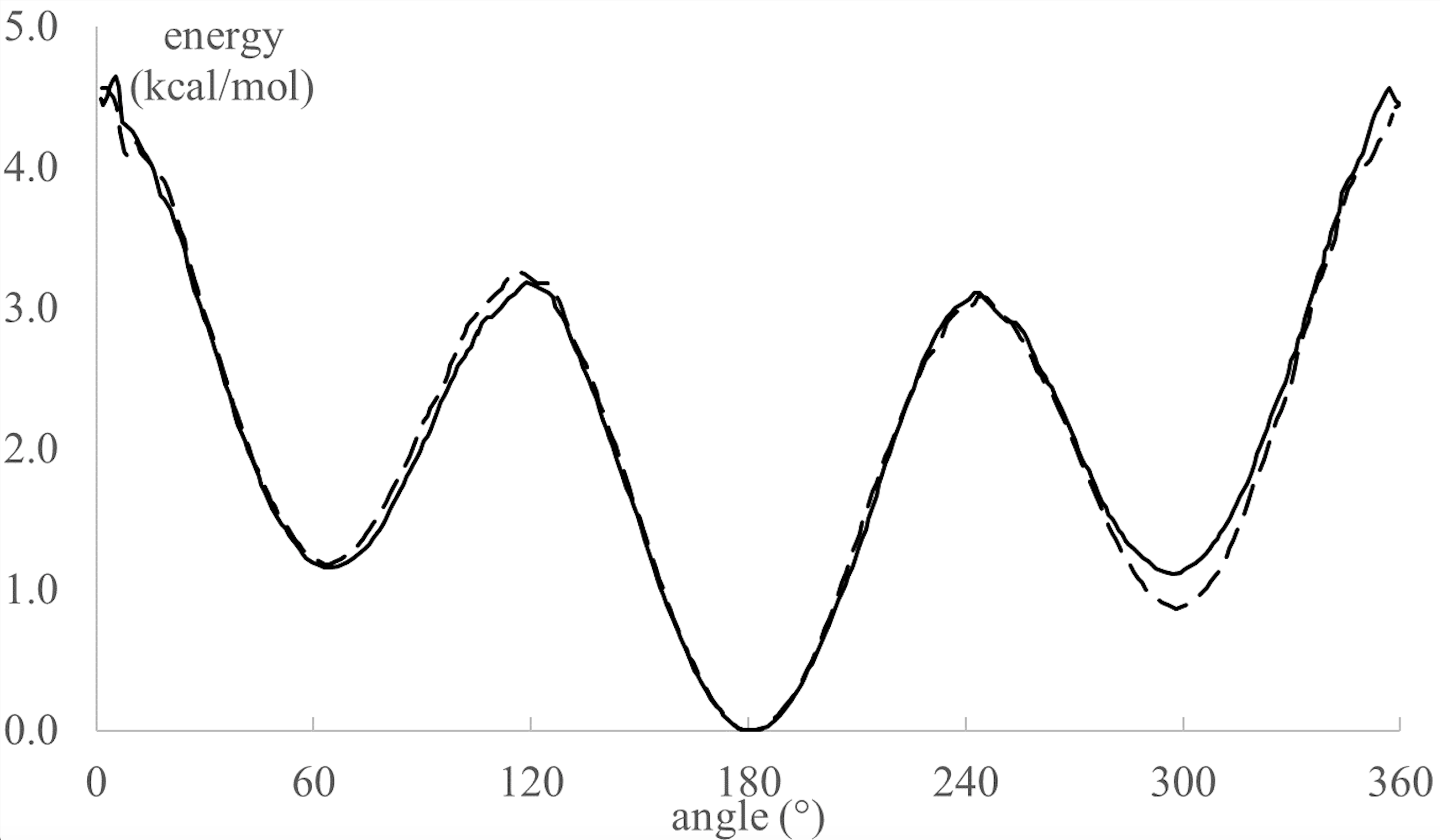}  
\caption{\small
Boltzmann-inverse analysis of an $\alkane {16}{32}$ (solid line), a $G_{16}$ molecule (dashed line) and the same $G_{16}$ molecule with the modified potential (dotted line). 
  {\bf Left Image:} bonds. {\bf Right Image:} valence angles.
  {\bf Middle Image:} dihedral angles. }
\label{boltzmann}
\end{center}
\end{myfigure}

One notices that the energies determined by Boltzmann-inverse for the molecules $\alkane {16}{32}$ and $G_{16}$ coincide very precisely. This result has been verified on longer molecules and at different temperatures.
As regards the modified potential, the results are unambiguous: it gives systematically higher energies which confirms that this potential is not the one we are looking for and that the dynamic mass is indeed $\gmol {14}$.

This second analysis at the local scale shows that the energies obtained by Boltzmann-inverse from AA and UA are well consistent.

\subsection{Comparison of End-to-End Distances}

Molecular conformational changes on a global scale depend mainly on dihedral angles,
moving from {\trans} conformations to {\gauche} (or {\gauchep}) conformations or vice-versa.
The {\gauche} conformations reduce the distance between the ends of the molecule,
called the end-to-end distance (\EtE).
One has analyzed the {\EtE} distances of $\alkane {16}{32}$ and $G_{16}$ molecules.
These distances fluctuate widely over the time due to the conformational changes. One lets temperatures vary from $\temp {100}$ to $\temp {2000}$
in steps of $\temp {50}$. The results are shown on \myfig {\ref {EtE}}.  \begin{myfigure} \begin{center}
  \includegraphics [width=9cm] {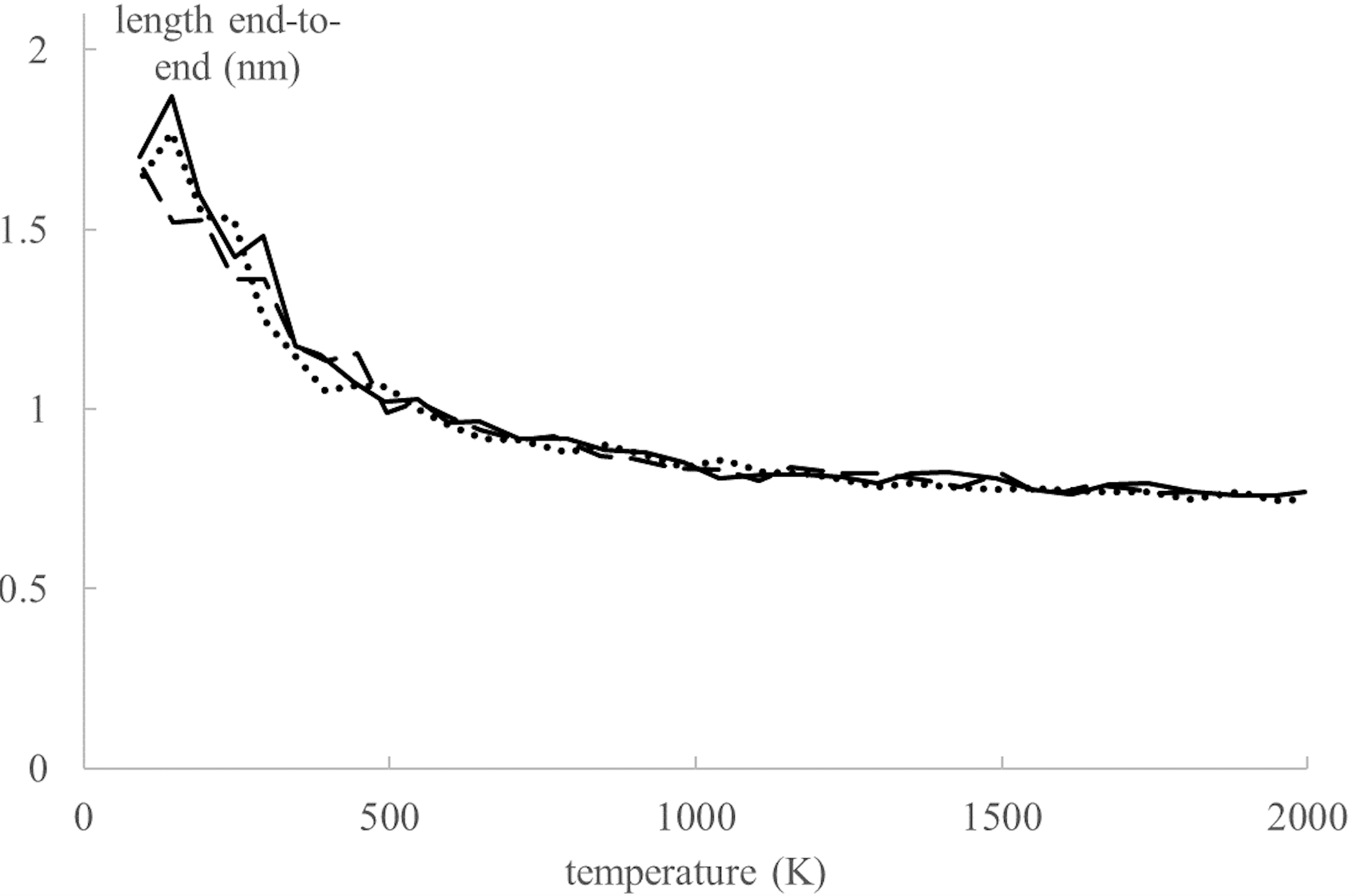}
  \caption{\small Distances $\EtE$ of $\alkane{16}{32}$ (solid line), $G_{16}$ (dashed line) and $G_{16}$ with the modified potential (dotted line).
}
\label{EtE}
\end{center}
\end{myfigure}
One can see that the average {\EtE} distances decrease when the temperature increases,
both in AA and UA, and that the results in UA are very close to the results in AA.

Very similar results are obtained using the modified potential introduced in \ref{vibration_period}.
which suggests the {\EtE} analysis is less sensitive to UA force field parameters than the local analysis.

Let us note that an {\EtE} analysis of a $C_{16}$ molecule was described in \cite {VAN_GUSTEREN} at a temperature of $\temp {373}$. The conclusion of this analysis, which differs from the one presented here, is that the {\EtE} distances in AA are greater than in UA.

\section {Conclusion \label {section:conclusion}}

A method was presented for deriving an UA potential from an AA potential based on an analysis of the correlations that may exist between coupled oscillators (e.g. CC and CH bonds).
The analyzed data are generated in molecular dynamics by simulations with durations in the nano-second range.
The method was applied to isolated alkanes, simulated from the OPLS-AA potential, and the intra-molecular components of the corresponding UA potential were determined.
The resulting bond stiffness constants and dihedral angle constants are the same as those in the article \cite{OPLS-UA} defining OPLS-AA.
We provide a justification, in terms of correlations, for the values of these constants.  However, the stiffness of the obtained UA valence angle potential differs from that of \cite {OPLS-UA} ($\kcalMolAngsquare {65.63}$ instead of $\kcalMolAngsquare {58.35}$).
Moreover, the UA potentials obtained do not depend on the size of the molecules, contrary to those of \cite{OPLS-UA}.

The method was validated by comparing the results of AA and UA simulations of similar molecules, at the same temperature, according to different criteria: mean end-to-end lengths of the molecules~;
mean periods of vibration of the oscillators~;
energies obtained by Boltzmann-inverse analysis,
based on the state probabilities of presence.
The comparisons made show the close proximity of the global kinematics of the AA and UA molecules.

The carbon atoms of all AA oscillators have the same dynamic mass of $\gmol {14}$, which means that the hydrogens are driven by the carbons to which they are bound.
Note that this does not necessarily imply that hydrocarboned oscillators are correlated to the carboned oscillators to which they are coupled.
Hence, there is no temporal correlation in the case of bonds, whereas the correlation is total in the case of dihedral angles, and only partial in the case of valence angles.

The correlation method seems to be easily applicable to molecules other than alkanes.




\end{document}